\documentclass[10pt,conference]{IEEEtran}
\IEEEoverridecommandlockouts

\usepackage{amsmath,amssymb,amsfonts}
\usepackage{algorithmic}
\usepackage{graphicx}
\usepackage{textcomp}
\usepackage{xcolor}
\usepackage[
bibstyle=ieee,
citestyle=numeric-comp,
maxnames=1,
minnames=1,
backend=biber,
]{biblatex}
\usepackage{hyperref}
\usepackage{cleveref}

\usepackage{siunitx}
\DeclareSIUnit{\Mbps}{Mbps}

\usepackage[absolute,overlay]{textpos} 
\usepackage{tcolorbox} 

\usepackage{tikz, tikzsymbols, pifont}
\usetikzlibrary{positioning, arrows.meta, calc, fit, matrix, patterns,
shapes.callouts, decorations.pathreplacing, shapes.geometric, shapes.misc,
shapes.symbols, automata}

\usepackage{bookmark}
\usepackage{multicol}

\usepackage[pro]{fontawesome5}

\usepackage{enumitem}

\graphicspath{{./images}}
\usepackage[normalem]{ulem}  

\usepackage{pbox}
\title{
{
    Evaluating UORA-Based Polling Mechanism for Latency-Sensitive Uplink Traffic
    in Wi-Fi Networks
{\footnotesize \textsuperscript{}
}}}

\author{\IEEEauthorblockN{Douglas Dziedzorm Agbeve, Andrey
    Belogaev, Chris Blondia, Jeroen Famaey}
\IEEEauthorblockA{
    \textit{IDLab, University of Antwerp – imec,
    Belgium}\\
\{douglas.agbeve, andrei.belogaev, chris.blondia, jeroen.famaey\}@uantwerpen.be}}

\begin{document}
\maketitle
\begin{textblock*}{\textwidth}(1.6cm, 25.7cm) 
    \centering
    \begin{tcolorbox}[colframe=black,
        colback=white,
        boxrule=0.4pt,
        width=0.98\textwidth,
         ] 

        \small \textcopyright~2025 IEEE. Personal use of this material is
        permitted. Permission from IEEE must be obtained for all other uses,
        in        any current or future media, including reprinting/republishing this
        material for advertising or promotional purposes, creating new
        collective works, for resale or redistribution to servers or lists, or
        reuse of any copyrighted component of this work in other works.
    \end{tcolorbox}
\end{textblock*}
\begin{abstract}
    IEEE 802.11ax (Wi-Fi 6) introduced Orthogonal Frequency Division Multiple
    Access (OFDMA), which enables simultaneous transmissions through centralized
    resource allocation. However, effective uplink scheduling requires the
    Access Point (AP) to identify which stations (STAs) have data to transmit. This
    typically necessitates polling for buffer status reports, a process that
    becomes increasingly inefficient and unscalable with growing device density.
    In this paper, we study how the Uplink OFDMA-based Random Access (UORA)
    feature improves the scalability and delay experienced by latency-sensitive
    data streams. We show that UORA enables efficient uplink scheduling while
    opportunistically identifying buffered traffic from unscheduled STAs,
    striking a balance between coordination and scalability. Performance
    evaluation of different polling strategies is done by means of simulation in
    ns-3. The results indicate that UORA-based polling outperforms alternative
    schemes in densely deployed network environments with heterogeneous uplink
    traffic patterns. Furthermore, under highly sparse and sporadic traffic
    conditions, UORA-based polling yields over \qty{40}{\percent} delay
    reduction compared to Scheduled Access (SA) OFDMA.
\end{abstract}


\section{Introduction \label{sec:intro}}

Enhanced Distributed Channel Access (EDCA) has traditionally served as the
primary channel access mechanism in Wi-Fi networks. This decentralized,
contention-based approach allows multiple stations (STAs) to independently
compete for the wireless medium. A STA that wins the contention gains access to
the entire channel bandwidth for its transmission. However, when multiple STAs
attempt to transmit simultaneously, collisions occur, resulting in packet loss
and necessitating retransmissions through the same contention process. EDCA
offers flexibility and does not have overhead related to transmission of
coordination and scheduling information from Access Point (AP) to STAs. However,
it lacks scalability and struggles to meet the Quality of Service (QoS)
requirements of latency-sensitive and high-reliability applications, such as
Extended Reality (XR) and Industrial Automation, especially in densely
deployed network environments \cite{adame2021time}. To address these challenges,
the IEEE 802.11ax amendment (Wi-Fi 6) introduced, among other features,
Orthogonal Frequency Division Multiple Access (OFDMA) as an alternative medium
access method to better manage the shared wireless spectrum~\cite{80211ax}.

OFDMA partitions the available bandwidth into multiple sub-bands called Resource
Units (RUs), which the AP can use to simultaneously schedule transmissions from
multiple STAs by assigning each a different RU. This allows multiple STAs to
transmit/receive data packets simultaneously thereby reducing contention and
collisions substantially. This structured resource allocation opens up new
possibilities for centralized resource management. In uplink (UL) OFDMA, for
instance, EDCA can be completely disabled on STAs, making the AP the sole
controller of all UL transmissions. However, to allocate RUs efficiently, the AP
must have knowledge of each STA's buffer status. This necessitates the
collection of Buffer Status Reports (BSRs) through UL polling mechanisms. As the
number of STAs increases, polling becomes progressively more challenging and
inefficient. Poorly timed and misdirected polling---such as assigning RUs to
STAs with empty buffers while those with data to send remain unserved---can lead
to significant resource underutilization and degraded network
performance~\cite{agbeve2025a2p}. To address this, Wi-Fi 6 introduces a hybrid
access mechanism known as Uplink OFDMA-based Random Access (UORA), which
combines the benefits of both scheduled and random access for buffer status
polling and data transmission.

During UORA transmissions, the AP divides the available RUs into Scheduled
Access (SA) and Random Access (RA) categories, and announces their allocation
via a Trigger Frame (TF). While scheduled STAs transmit without contention using
their assigned SA RUs, all other STAs may contend for RA RUs using an UORA
random access procedure (cf., Section~\ref{sec:bckgrnd}). This hybrid
approach enables efficient scheduling of known buffered data while
simultaneously supporting scalable, opportunistic polling for the unknown buffer
status of STAs, thereby balancing centralized coordination with adaptive
responsiveness. In this paper, we study specifically the performance of UORA as
polling mechanism.

The main contributions of this paper are as follows:
\begin{itemize}
    \item We evaluate the efficacy of UORA as a scalable alternative to buffer
        status polling mechanisms such as A2P~\cite{agbeve2025a2p} and SA UL
        OFDMA, hereafter also referred to as SA OFDMA.
    \item We investigate the impact of the minimum OFDMA contention window
        size on the performance of UORA under varying traffic loads.
    \item We demonstrate that, beyond a certain threshold of traffic sparsity,
        the performance advantage of UORA over SA OFDMA plateaus.
\end{itemize}

The remainder of the paper is organized as follows. Section~\ref{related_works}
reviews recent research that addresses the polling problem and considers UORA in
OFDMA-based Wi-Fi networks. Section~\ref{sec:bckgrnd} presents an overview of
key OFDMA parameters and the UORA procedure. We discuss the relevant
implementation details of the various schemes in Section~\ref{sec:method}. In
Section~\ref{evaluation}, we conduct a comparative performance evaluation of
UORA against alternative approaches. Finally, Section~\ref{conclusion} concludes
the paper.

\section{Related Work \label{related_works}}

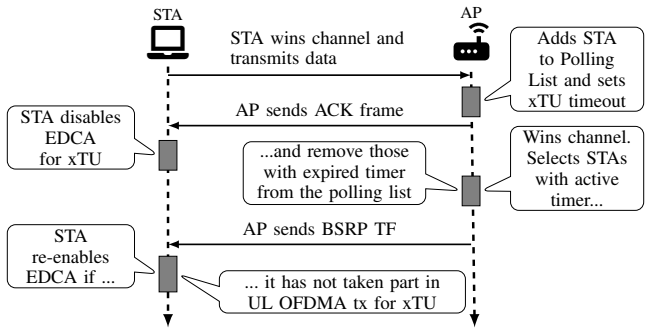
\begin{figure}[tb]
    \centering
    \scalebox{0.79}{
    \begin{tikzpicture}

        \node (STA) at (-5,4) {\scalebox{1.6}{\faLaptop}};
        \node (AP) at (.1,4) {\scalebox{.21}{\includegraphics[trim=0 0 0 0,
            clip]{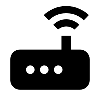}}};

        \node[above, font=\footnotesize, yshift=-.15cm] at (STA.north) {STA};
        \node[above, font=\footnotesize, yshift=-.2cm] at (AP.north) {AP};

        \draw[-{latex}, dashed, very thick] (STA.south) -- (-5,-.8);
        \draw[-{latex}, dashed, very thick] (AP.south) -- (.16,-.8);

            \draw[-{latex[length=12mm, width=10mm]}, thick] (-5, 3.45) --
                (.1,3.45)
                node[midway, above, font=\footnotesize] (desc) {};
            \node[ font=\small, inner sep=1pt, rounded corners=2pt,
                yshift=.2cm, text width=3cm] at (desc.north) {STA wins channel
                and transmits data};

            \node[draw, minimum height=.5cm,  minimum
                width=.3cm, fill=black!50] (addtoList) at (.1,3) {};

            \node[draw, rectangle callout, callout absolute
                pointer=(addtoList.east), rounded corners, text width=2.1cm,
                font=\small, right= of addtoList.east, xshift=-.5cm,
                yshift=.55cm, inner sep=2pt, align=center,
                ] (FIRSTCALLOUT1){Adds STA to Polling List and sets xTU
                timeout};

            \draw[-{latex[width=5mm]}, thick] (.1, 2.6) -- (-5,
                2.6)
                node[midway, above, font=\footnotesize] (desc1) {};
            \node[font=\small, inner sep=1pt, rounded corners=2pt,
                yshift=0cm] at (desc1.north) {AP sends ACK frame};

            \node[draw, minimum height=.5cm,  minimum
                width=.3cm, fill=black!50] (addtoList1) at (-5, 2.1) {};

            \node[draw, rectangle callout, callout absolute
                pointer=(addtoList1.west), rounded corners, text width=2cm,
                font=\small, left= of addtoList1.west, xshift=.6cm,
                yshift=.3cm, inner sep=2pt, align=center,
                ] (FIRSTCALLOUT2){STA disables EDCA for xTU};

            \node[draw, minimum height=.5cm,  minimum
                width=.3cm, fill=black!50] (addtoList2) at (.1,1.5) {};

            \node[draw, rectangle callout, callout absolute
                pointer=(addtoList2.east), rounded corners, text width=2.1cm,
                font=\small, right= of addtoList2.east, xshift=-.5cm,
                yshift=.38cm, inner sep=2pt, align=center,
                ] (FIRSTCALLOUT3){Wins channel. Selects STAs with active
                timer...};
            \node[draw, rectangle callout, callout absolute
                pointer=(addtoList2.west), rounded corners, text width=3cm,
                font=\small, left= of addtoList2.west, xshift=.4cm,
                yshift=.3cm, inner sep=1pt, align=center,
                ] (FIRSTCALLOUT4){...and remove those with expired timer from
                the polling list};

            \draw[-{latex[width=5mm]}, thick] (.1,0.6) -- (-5, 0.6)
                node[midway, above, font=\footnotesize] (desc1) {};
            \node[font=\small, inner sep=1pt, rounded corners=2pt,
                yshift=0cm] at (desc1.north) {AP sends BSRP TF };

            \node[draw, minimum height=.6cm, minimum
                width=.3cm, fill=black!50] (addtoList3) at (-5, 0.1) {};

            \node[draw, rectangle callout, callout absolute
                pointer=(addtoList3.west), rounded corners, text width=2cm,
                font=\small, left= of addtoList3.west, xshift=.6cm,
                yshift=.3cm, inner sep=2pt, align=center,
                ] (FIRSTCALLOUT2){STA \\re-enables EDCA  if ...};

            \node[draw, rectangle callout, callout absolute
                pointer=(addtoList3.east), rounded corners, text width=4cm,
                font=\small, right= of addtoList3.east, xshift=-.3cm,
                yshift=-.3cm, inner sep=2pt, align=center,
                ] (FIRSTCALLOUT2){... it has not taken part in UL OFDMA tx for
                xTU};

\end{tikzpicture}
}
\caption{\footnotesize{Diagramatic representation of the A2P
algorithm~\cite{agbeve2025a2p}}}
\label{fig:a2p_algo}
\vspace{-.4cm}
\end{figure}

A considerable number of studies in the literature have explored strategies to
enhance OFDMA performance in Wi-Fi networks, with a strong emphasis on resource
allocation mechanisms~\cite{wang2018scheduling, bankov2018ofdma,
inamullah2020will, qadri2022preparing, noh2024joint, tan2024deep}. However, the
role of BSR collection---an essential prerequisite for effective uplink
scheduling---is often underexplored. For instance, the resource allocation task
has been framed as an optimization problem and addressed using a sub-optimal
divide-and-conquer recursive algorithm \cite{wang2018scheduling}. Classical
scheduling techniques—MaxRate, Proportional Fair, and Shortest Remaining
Processing Time—have been adapted for Wi-Fi \cite{bankov2018ofdma}. A scheduling
algorithm for deadline-aware traffic has also been proposed, utilizing queuing
theory on buffer state information reported by STAs to estimate the Head-of-Line
(HOL) delay \cite{inamullah2020will}. A related heuristic prioritizes
transmissions based on buffer sizes across different access categories
\cite{qadri2022preparing}. More recent efforts apply deep reinforcement learning
to improve scheduling efficiency \cite{noh2024joint, tan2024deep}. While buffer
status information is central to many of these approaches, the overhead and
challenges associated with BSR collection are generally overlooked. To address
these limitations, we proposed the A2P algorithm in our earlier
paper~\cite{agbeve2025a2p}. A2P improves resource utilization efficiency by
combining EDCA and OFDMA. The AP tracks a polling list of STAs expected to
transmit, allowing them to use only UL OFDMA and bypass EDCA contention. A STA
joins the list by initiating a transmission via EDCA. After successful
reception, the AP disables EDCA for that STA for a preconfigured time period
using the MU EDCA Parameter Set. STAs that have not reported any data after
being polled during the pre-specified time period are removed from the polling
list. Figure~\ref{fig:a2p_algo} illustrates the mechanism.

As it operates on a random access mechanism, UORA is susceptible to collisions
and the subsequent need for retransmissions of lost packets. For this reason, it
is most suited for BSRs, which are significantly smaller than full data packets,
thereby limiting the impact of collision overhead. Several proposals in the
literature seek to improve the performance of UORA. For example, researchers
have explored more efficient ways of selecting the $\text{OBO}$
counter~\cite{rehman2023collision, kim2021ofdma, liu2025performance}, optimizing
the back-off countdown procedure~\cite{kosek2022efficient, kosek2022improving,
rehman2025enhancing}, improving resource allocation~\cite{bhattarai2019uplink}
and addressing collision resolution~\cite{avdotin2019ofdma, avdotin2019enabling,
xie2020multi}. Other works have integrated UORA with recently introduced Wi-Fi
features, such as multi-link operation~\cite{jin2024enhancing}, and target wake
time~\cite{chen2020scheduling}. Furthermore, some studies have proposed
scheduling algorithms that combine both random and scheduled access
procedures~\cite{bhattarai2019uplink, avdotin2020resource}. The performance of
UORA has also been evaluated using analytical modeling and simulation
\cite{naik2018performace}. However, their publicly available UORA
implementation in ns-3 exhibits several limitations, which we addressed in our
work~\cite{agbeve2025design}. Our implementation is available as open
source~\cite{UORAns3}.

Several alternative solutions to UORA have been proposed in the literature. A
fully deterministic channel access method has been introduced, in which the AP
centrally schedules all uplink transmissions by disabling random access
altogether \cite{schneider2023deterministic}. Although this method is effective
for predictable traffic patterns, it tends to perform poorly under bursty or
unpredictable traffic conditions. Another approach involves using multiple
rounds of BSRP Trigger Frames to collect BSRs from all STAs prior to scheduling,
which enhances fairness but increases the overhead associated with BSR
collection \cite{shao2024access}. Additionally, a mechanism that enables
client-side switching between EDCA and OFDMA based on buffer status has been
proposed \cite{goncalves2023access}. However, this conflicts with the standard,
where the AP governs access states, and bypassing the AP risks inconsistent
behavior and inefficient resource use. To the best of our knowledge, no prior
work has evaluated the effectiveness of using UORA as a polling mechanism in
comparison with other approaches such as SA UL OFDMA.

In this paper, we assess the performance of UORA-based polling for buffer status
reporting and compare it against alternative mechanisms, including SA UL OFDMA and
A2P. EDCA serves as the baseline approach.

\section{Background: SA OFDMA and UORA in Wi-Fi \label{sec:bckgrnd}}

In this section, we detail the general principles of SA OFDMA, followed by a
discussion of the MU EDCA Parameter Set, and conclude with the operational
specifics of UORA.

\subsection{SA UL OFDMA}
OFDMA, debuted in the Wi-Fi 6 standard, enables simultaneous frame transmissions
to and from STAs by dividing the available bandwidth into multiple RUs, which
can be allocated to different STAs. The standard supports both downlink (DL) and
UL OFDMA transmissions. However, as this work focuses on UORA, which is an UL
transmission mechanism, DL procedures are beyond the scope of this discussion.

Once the AP successfully gains channel access through contention, it initiates
UL OFDMA transmission by broadcasting a Trigger Frame (TF), specifically a Basic
TF. This frame includes, among other parameters, the mappings of selected STAs
to their assigned RUs. Following a Short Inter-Frame Space (SIFS), the scheduled
STAs transmit concurrently using their allocated RUs. Upon completion of the UL
transmission, the AP responds, following another SIFS, with a Multi-STA Block
Acknowledgement (BA) to confirm successful packet reception. To allocate RUs
effectively for UL transmissions, the AP must first determine which STAs have
buffered data. This is achieved by sending a special type of TF, known as a
Buffer Status Report Poll (BSRP). Similar to a Basic TF, the BSRP frame includes
RU assignments that indicate which STAs should respond. STAs transmit their buffer
status reports (BSRs) either explicitly by sending a QoS Null frame--triggered
when the user index in the BSRP TF matches their Association ID (AID)--or
implicitly by embedding the buffer size in the QoS control field of any outgoing
frame. Based on these reports, the AP identifies STAs with pending data and
considers them for scheduling in subsequent UL transmissions. To enhance airtime
fairness and minimize contention in dense deployments, the standard introduced
the Multi-User (MU) EDCA Parameter Set. STAs participating in UL OFDMA
transmissions apply this set of parameters to contend for the channel less
aggressively or defer access entirely for a specified duration.
Figure~\ref{fig:ul_ofdma} illustrates the complete SA UL OFDMA frame exchange
sequence.
\tikzstyle{sta_freq} = [rectangle, draw, align=flush center, thin,
                        minimum height = .5cm, minimum width = 1cm, outer sep =
                        0pt]
\tikzstyle{ap_freq} = [rectangle, draw, align=flush center, thin,
                        minimum height = 2.5cm, minimum width = 1.0cm]
\tikzstyle{line} = [-{latex[length=7mm, width=5mm]}, thin,
                        draw,color=black!100]
\tikzstyle{sifs} = [{latex[length=7mm, width=5mm]}-{latex[length=7mm,
width=5mm]}, thin, draw,color=black!100]
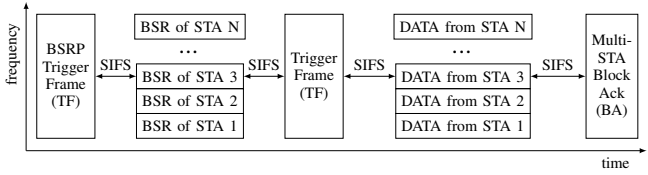
\begin{figure}[t]
    \centering
        \scalebox{0.66}{
            \centering
            \begin{tikzpicture}[node distance=2.5cm, font=\small, on
                grid, auto, label position=center, align = flush center]
                \node[ap_freq](AP1) {BSRP\\Trigger\\Frame\\(TF)};
                \node[sta_freq, right of = AP1](STA1){BSR of STA 3};
                \node[sta_freq, below of = STA1, node distance=.5cm](STA2){BSR
                    of STA 2};
                \node[sta_freq, below of = STA2, node distance=.5cm](STA3){BSR
                    of STA 1};
                \node[above of = STA1, node distance =
                    .5cm](STA4){\textbf{\dots}};
                \node[sta_freq, above of = STA4, node distance=.5cm](STA5){BSR
                    of STA N};
                \node[ap_freq, right of = STA1](AP2)
                    {Trigger\\Frame\\(TF)};
                 \node[sta_freq, right of = AP2, node distance =
                     3cm](STA6){DATA from  STA 3};
                \node[sta_freq, below of = STA6, node
                    distance=.5cm](STA7){DATA from STA 2};
                \node[sta_freq, below of = STA7, node
                    distance=.5cm](STA8){DATA from STA 1};
                \node[above of = STA6, node distance =
                    .5cm](STA9){\textbf{\dots}};
                \node[sta_freq, above of = STA9, node
                    distance=.5cm](STA10){DATA from STA N};
                \node[ap_freq, right of = STA6, node distance = 3cm](AP3)
                    {Multi-\\STA\\Block\\Ack\\(BA)};

                \path[line](-.8,-1.5) -- node[pos=.65, rotate=90,
                    anchor=south]{frequency}++(0,3);
                \path[line](-.8,-1.5) -- node[pos=0.95,
                    anchor=north]{time}++(12.5,0);
                \path[sifs](AP1)--node[pos=0.5]{SIFS}(STA1);
                \path[sifs](STA1)--node[pos=0.5]{SIFS}(AP2);
                \path[sifs](AP2)--node[pos=0.5]{SIFS}(STA6);
                \path[sifs](STA6)--node[pos=0.5]{SIFS}(AP3);
            \end{tikzpicture}
        }
        \vspace{-.7cm}
    \caption{UL OFDMA frame exchange sequence}
    \label{fig:ul_ofdma}
    \vspace{-.3cm}
\end{figure}

\subsection{MU EDCA Parameter Set}
The MU EDCA Parameter Set includes EDCA parameters such as contention window and
Arbitration Inter-Frame Space Number (AIFSN) that can be used by STAs after
participating in UL OFDMA transmission. By using it, the AP can exert greater
control over UL transmissions, while the STAs themselves compete less
aggressively for the channel or do not compete at all. The AP announces the
MU-EDCA Parameter Set through management frames. When the parameter set includes
an AIFSN value of zero, it signals the STAs to completely disable EDCA-based
contention. In such cases, the AP fully orchestrates UL transmissions, and STAs
do not contend for medium access. In addition to EDCA parameters, the MU EDCA
Parameter Set includes an MU EDCA timer that dictates how long a STA should
apply the received parameters. The timer is reset each time the STA successfully
transmits data via OFDMA and receives a corresponding Block Ack from the AP. If
the timer expires without a successful transmission, the STA reverts to its
default EDCA settings.

\subsection{The UORA operation}
\tikzstyle{sta_freq} = [rectangle, draw, align=flush center, thin,
                        minimum height = .5cm, minimum width = 2.3cm, outer sep
                        = 0pt]
\tikzstyle{ap_freq} = [rectangle, draw, align=flush center, thin,
                        minimum height = .5cm, minimum width = 1cm]
\tikzstyle{line} = [-{latex[length=7mm, width=5mm]}, thin,
                        draw,color=black!100]
\tikzstyle{sifs} = [{latex[length=7mm, width=5mm]}-{latex[length=7mm,
width=5mm]}, thin, draw,color=black!100]

\begin{figure}[t]
    \centering
        \scalebox{0.635}{
            \begin{tikzpicture}[node distance = .7cm, on
                grid, auto, label position=center, align = flush center]
                \node[ap_freq, xshift = -1.2cm, font=\small](BSRP_TF1) {AID 0};
                \node at (BSRP_TF1.north) [above, font=\small ] {BSRP TF};
                \node[ap_freq, below = of BSRP_TF1, yshift = .2cm,
                    font=\small](BSRP_TF2) {AID 0};
                \node[ap_freq, below = of BSRP_TF2, yshift = .2cm,
                    font=\small](BSRP_TF3) {AID 0};
                \node[ap_freq, below = of BSRP_TF3, yshift = .2cm,
                    font=\small](BSRP_TF4) {AID 5};
                \node[ap_freq, below = of BSRP_TF4, yshift = .2cm,
                    font=\small](BSRP_TF5) {AID 7};
                \node[sta_freq, right = of BSRP_TF1.east, xshift =0cm,
                    fill=gray!50, font=\small](BSR_STA1){BSR STAs 2,8};
                \node[sta_freq, below = of BSR_STA1, yshift =
                    .2cm, font=\small](BSR_STA2){Idle RU};
                \node[sta_freq, below = of BSR_STA2, yshift = .2cm,
                    font=\small](BSR_STA3){BSR STA 3};
                \node[sta_freq, below = of BSR_STA3, yshift = .2cm,
                    font=\small](BSR_STA4){BSR STA 5};
                 \node[sta_freq, below = of  BSR_STA4, yshift = .2cm,
                     font=\small](BSR_STA5){BSR STA 7};
                \node[ap_freq, right = of BSR_STA1.east, minimum height = 2.5cm,
                    yshift = -1cm, font=\small](BlockACK1)
                    {Multi-\\STA\\Block\\Ack\\(BA)};
                \node[ap_freq, right = of BlockACK1.east, yshift =
                    1cm, font=\small](BASIC_TF1) {AID 6};
                \node at (BASIC_TF1.north) [above, font=\small] {Basic TF};
                \node[ap_freq, below = of BASIC_TF1, yshift = .2cm,
                    font=\small](BASIC_TF2) {AID 1};
                \node[ap_freq, below = of BASIC_TF2, yshift = .2cm,
                    font=\small](BASIC_TF3) {AID 3};
                \node[ap_freq, below = of BASIC_TF3, yshift = .2cm,
                    font=\small](BASIC_TF4) {AID 5};
                \node[ap_freq, below = of BASIC_TF4, yshift = .2cm,
                    font=\small](BASIC_TF5) {AID 7};
                 \node[sta_freq, right = of BASIC_TF1.east,
                     font=\small](DATA_STA1){DATA STA 6};
                \node[sta_freq, below = of DATA_STA1, yshift = .2cm, font=\small
                    ](DATA_STA2){DATA STA 1};
                \node[sta_freq, below = of DATA_STA2, yshift = .2cm, font=\small
                    ](DATA_STA3){DATA STA 3};
                \node[sta_freq, below = of DATA_STA3, yshift = .2cm, font=\small
                    ](DATA_STA4){DATA STA 5};
                \node[sta_freq, below = of DATA_STA4, yshift = .2cm, font=\small
                    ](DATA_STA5){DATA STA 7};
                \node[ap_freq, right = of DATA_STA1.east, minimum height = 2.5cm,
                    yshift = -1cm, font=\small](BlockACK2)
                    {Multi-\\STA\\Block\\Ack\\(BA)};
                \path[line](-2,-2.5) -- node[pos=.5, rotate=90,
                    anchor=south, font=\small]{frequency}++(0,3.5);
                \path[line](-2,-2.5) -- node[pos=0.95,
                    anchor=north, font=\small]{time}++(13,0);
                \path[sifs](BSRP_TF3.east)--node[pos=0.5,
                    font=\small]{SIFS}(BSR_STA3.west);
                \path[sifs](BSR_STA3.east)--node[pos=0.5,
                    font=\small]{SIFS}(BlockACK1.west);
                \path[sifs](BlockACK1.east)--node[pos=0.5,
                    font=\small]{SIFS}(BASIC_TF3.west);
                \path[sifs](BASIC_TF3.east)--node[pos=0.5,
                    font=\small]{SIFS}(DATA_STA3.west);
                \path[sifs](DATA_STA3.east)--node[pos=0.5,
                    font=\small]{SIFS}(BlockACK2.west);
                \node at (-1, -5) (first_table) [font=\small] {
                        \renewcommand{\arraystretch}{1.2}
                        \begin{tabular}{|c|c|}
                            \hline
                            STA & OBO \\
                            \hline
                            1 & 15 \\
                            \hline
                            2 & 1  \\
                            \hline
                            3 & 2 \\
                            \hline
                            4 & 5 \\
                            \hline
                            6  & 7 \\
                            \hline
                            8 & 3\\
                            \hline
                        \end{tabular}
                    };
                \node at (first_table.north) [above, font=\small ] {Before BSRP
                    TF};
                \node [right = of first_table.east, font=\small] (first_table) {
                        \renewcommand{\arraystretch}{1.2}
                        \begin{tabular}{|c|c|}
                            \hline
                            STA & OBO \\
                            \hline
                            1 & 12 \\
                            \hline
                            2 & 0  \\
                            \hline
                            3 & 0 \\
                            \hline
                            4 & 2 \\
                            \hline
                            6 & 4 \\
                            \hline
                            8 & 0\\
                            \hline
                        \end{tabular}
                    };
                \node at (first_table.north) [above, font=\small ] {After BSRP
                    TF};
                \draw[red, thick] (-2.05cm, -4.35cm) rectangle (3.34cm,-4.8cm);
                \draw[red, thick] (-2.05cm, -5.2cm) rectangle (3.34cm,-4.8cm);
                \draw[red, thick] (-2.05cm, -6.54cm) rectangle (3.34cm,-6.1cm);
            \end{tikzpicture}
        }
    \caption{UL OFDMA frame exchange sequence with UORA. Tables in the bottom
        show the values of OFDMA Back-Off (OBO) for all STAs before and after
        BSRP TF is received. STAs whose OBO reaches 0 are framed with red.
    }
    \label{fig:uora_ofdma}
    \vspace{-.3cm}
\end{figure}
As with all UL transmissions orchestrated by the AP, the AP initiates the
transmission by broadcasting a TF to signal its start. In UORA, the AP can
designate a subset of RUs in the TF for either RA, allowing all STAs to use them,
or allocate an RU for SA, restricting its use to a single designated STA. During
the association stage, the AP shares information about the OFDMA Contention
Window ($\text{OCW}$) range defined by $\text{EOCW}_{\text{min}}$ and
$\text{EOCW}_{\text{max}}$. These parameters are transmitted in the management
frames, and their values can be adjusted on demand. If a STA receives a TF that
does not explicitly assign it an RU, but the frame indicates that RA is
permitted, the STA may attempt UL transmission via UORA---for instance, to
transmit a new buffer status report. In this scenario, the STA initializes its
$\text{OCW}$ to $\text{OCW}_{\text{MIN}} = \text{2}^{\text{EOCW}_{\text{min}}} -
\text{1}$ and randomly selects an initial OFDMA Back-Off ($\text{OBO}$) value
within the range $[\text{0}, \text{OCW}]$. Upon receiving subsequent TFs, the
STA decreases its $\text{OBO}$ value by the number of RA RUs specified in the
TF. If the updated $\text{OBO}$ counter is less than or equal to the number of
RA RUs, the STA randomly chooses one of advertised RA RUs in the TF and uses it
to transmit. Following a successful transmission, the STA resets its
$\text{OCW}$ to $\text{OCW}_{\text{MIN}}$. However, if transmission fails (e.g.,
due to collision), the STA doubles its $\text{OCW}$ up to an upper bound of
$\text{OCW}_{\text{MAX}} = \text{2}^{\text{EOCW}_{\text{max}}} - \text{1}$.

Figure~\ref{fig:uora_ofdma} depicts the UORA frame exchange sequence. The AP
initiates an UL OFDMA transmission by sending a BSRP TF, which includes three RA
RUs (denoted by AID 0) and two SA RUs assigned to STAs 5 and 7. This BSRP TF
prompts STAs to report their buffer status, enabling the AP to identify which
STAs require resources. After one SIFS, STAs 2, 3, 4, 7 and 8 transmit BSRs
using either a randomly selected RA RU or an SA RU assigned to them. STAs 2, 3,
and 8 are eligible to transmit in a RA RU, because their $\text{OBO}$ values are
less than or equal to 3, the number of RA RUs advertised in the BSRP TF.
Specifically, STA 3 selects the third RA RU and successfully transmits, while
STAs 2 and 8 transmit using the same RA RU, resulting in a collision (shown as
the shaded area). Following this, the AP acknowledges the transmission by
sending a Multi-STA Block ACK after a SIFS. STAs with unsuccessful transmissions
double their $\text{OCW}$ values and choose new $\text{OBO}$ values. After
another SIFS, the AP allocates RUs to STAs that have reported having data to
transmit (STAs 3, 5, and 7). The AP can also allocate resources to STAs that
have previously reported non-zero buffer statuses, or to STAs that it thinks
might have data for transmission but failed to deliver their buffer statuses.
For example, in the figure, the AP also allocates resources to STAs 1 and 6. It
is also allowed to assign some of the RUs for random access, but the overhead
due to collisions of the data packets is usually significantly higher than that
for BSRs. This is because the transmission time of data packets is generally
longer than that of BSRs. Note that RUs can be of different sizes depending on
the needs of the STAs, but for simplicity, they are considered the same in the
figure. This allocation is communicated through the Basic TF. Following one more
SIFS, the STAs transmit on their respective RUs. To ensure synchronized
transmission, smaller payloads are padded to match the size of the largest
payload. Finally, the AP sends a Multi-STA Block ACK after a SIFS, thereby
concluding the UL OFDMA transmission.

\section{Implementation \label{sec:method}}

In this section, we discus the relevant implementation details of UORA, SA
OFDMA, A2P, and EDCA. The operational differences among the various schemes
are:
\begin{enumerate}
    \item \emph{UORA:} A subset of RUs is reserved for RA in the TF, allowing
        unscheduled STAs to transmit buffer status reports opportunistically.
    \item \emph{SA OFDMA:} All RUs are allocated for SA without reserving any
        RUs for RA.
    \item \emph{A2P:} STAs indicate their need for resources by transmitting
        their initial packet via the contention-based EDCA mechanism, thereby
        prompting the AP to schedule them in subsequent transmissions.
    \item \emph{EDCA:} All STAs contend for the channel using traditional
        EDCA, transmitting over the entire bandwidth upon winning access.
\end{enumerate}

For the OFDMA-based mechanisms (UORA, SA OFDMA, and A2P), multi-user
transmissions are intentionally delayed on both the AP and STAs for a brief
period following system initialization. This delay ensures that initial setup
procedures---such as establishing acknowledgments---are completed via EDCA,
which remains uninterrupted by the absence of multi-user transmissions during 
this initialization phase. Additionally, the AP is configured to request channel
access even when it does not have data queued for transmission. This
configuration is necessary because our experiments do not include downlink
traffic, and thus the AP would otherwise have no opportunity to contend for
channel access. The time between consecutive access requests is referred to as
the Access Request Interval (ARI).

In SA OFDMA and A2P, the AP allocates all available RUs to selected STAs for
both buffer status reporting and data transmission in a round-robin manner.
However, under UORA, only the RUs reserved for SA are assigned in round-robin
fashion for BSRs, while all available RUs are scheduled in the Basic TF for data
transmission. This implies that, RUs left unused during the BSRP/BSR
exchange---either due to collision or lack of selection---are subsequently
scheduled for data transmission. In A2P, the AP selects STAs from a polling
list, whereas in SA OFDMA and UORA, all associated STAs are considered.

\section{Performance Evaluation \label{evaluation}}
We evaluate the performance of different schemes using the ns-3 network
simulator~\cite{henderson2008ns3}. In particular, we used our previously
developed open-source UORA implementation~\cite{agbeve2025design, UORAns3} for
assessment and adapt the A2P algorithm~\cite{agbeve2025a2p} to suit the targeted
traffic pattern. We compare the performance of the following four schemes; UORA,
SA OFDMA, A2P and EDCA.

The comparative evaluation focuses on the \emph{Uplink Delay} of the
latency-sensitive STAs and the \emph{Total Throughput} of all the associated
STAs. \emph{Uplink Delay} is defined as the time interval between the generation
of a packet by a STA and its successful reception by the AP. \emph{Total
Throughput} refers to the aggregate rate, measured in packets per second, of
packets successfully received by the AP from all STAs during the simulation.

\subsection{Simulation Setup}
In the experiments, we model a single Basic Service Set (BSS) consisting of
multiple STAs and a single AP, both compliant with the Wi-Fi 6 (or newer) standard.
The STAs are categorized into two groups, \emph{deterministic} and
\emph{stochastic}, based on their traffic generation behavior. Deterministic
STAs generate Constant Bit Rate (CBR) UDP  traffic at approximately
\qty{6.54}{\Mbps} (i.e., $(1700 \times 8)~\text{bits} \div 0.00208~\text{s}$ ),
while stochastic STAs generate latency-sensitive UDP traffic with a fixed packet
size of \qty{1700}{\byte} following an exponentially distributed packet arrival
rate. All traffic is assigned to
the Voice (VO) Access Category (AC). This configuration emulates a network of
wirelessly connected devices sharing the same AC, where some devices generate
data at regular intervals, while others generate event-based latency-sensitive
traffic unpredictably.

We set packet size to 1700 bytes to ensure that each transmission fits entirely
within the allocated Transmit Opportunity (TXOP) for data transmission,
particularly in the OFDMA based (i.e., SA OFDMA, UORA and A2P) scenarios where
only 26-tone RUs are employed. In the A2P configuration, frame aggregation is
disabled to avoid skewing the expected performance outcome, as it could allow
multiple packets to be bundled into the initial EDCA transmission, thereby
masking the intended behavior. Conversely, aggregation is enabled in the
EDCA simulation to bolster performance. We select the smallest RU type
(i.e., 26-tones) to promote equitable distribution of resources among associated
STAs and maximize the number of simultaneous transmissions. The number of
deterministic STAs is equal to the number of 26-tone RUs in the chosen
bandwidth. Additionally, downlink traffic generation is deliberately disabled,
as this study focuses solely on the polling mechanisms used for UL
transmissions.

In the UORA setup, EDCA is fully disabled by setting the MU EDCA timer to the
full simulation duration and the AIFSN value to zero. This configuration
delegates full control of UL transmission scheduling to the AP, with STAs
refraining from any independent contention. Accordingly, the ARI is set to one
Short Inter-Frame Space (SIFS) to enable frequent polling and UL data
transmissions. In contrast, the A2P configuration allows the AP to request
access less aggressively, enabling fairer contention opportunities for STAs that
are not on the polling list. To avoid saturating the polling list with
non-transmitting stochastic STAs, the MU EDCA timer is configured to
\num{8} Time Units (TUs), which is the minimum duration allowed by the
standard. Furthermore, the $\text{OCW}_\text{MIN}$ value is varied across
simulations to evaluate its effect on performance, while $\text{OCW}_\text{MAX}$
is held constant at its maximum value.

To ensure that packet loss due to channel errors is negligible, we configure
transmissions to be at sufficiently high power levels such that all STAs remain
within the communication range of the AP. Consequently, packet losses only occur
when multiple STAs transmit at the same time using the same resource (RU or
bandwidth). The simulation parameters used across all scenarios are summarized
in Table~\ref{table:sim_para}.

\begin{table}[tb]
    \caption{List of Simulation Parameters}
    \vspace{-10pt}
    \label{table:sim_para}
    \begin{center}
        {\renewcommand{\arraystretch}{1.8}%
        \begin{tabular}{l c}
            \hline
            \textbf{Parameter} & \textbf{Value} \\
            \hline
            Carrier frequency & \qty{5}{\giga\hertz} \\
            Bandwidth & \qty{20}{\mega\hertz} \\
            Guard Interval & \qty{0.8}{\micro\second} \\
            MCS Index & \num{8} \\
            Resource Unit Type & 26\hyphen tone only\\
            Transmit Opportunity & \qty{2.08}{\milli\second} \vspace{3pt}\\
            AP Access Request Interval (ARI) &
            \pbox{2cm}{\qty{16}{\micro\second}, UORA\\
            \qty{128}{\micro\second}, A2P} \\
            EDCA Access Category & VO \\
            $\text{OCW}_\text{min}$ & \{0, 1, 3, 7, 15, 31, 63\} \\
            $\text{OCW}_\text{max}$ & 127 \vspace{3pt}\\
            MU EDCA Timer & \pbox{2cm}{\qty{180}{\second}, UORA
            \\\qty{8}{TUs}, A2P} \vspace{3pt}\\
            Payload Size & \qty{1700}{\byte} \\
            Deterministic Inter-Packet Interval & \qty{2.08}{\milli\second}
            \vspace{3pt} \\
                Stochastic Inter-Packet Interval  &
                \pbox{8cm}{Exp. Distribution, \\(means: \{0.03, 0.05, \\0.1,
                0.3, 0.5, 1.0\} s)} \\
            Number of Deterministic STAs & \num{9} \\
            Duration of Simulation & \qty{180}{\second} \\
            \hline
        \end{tabular} }
    \end{center}
    \vspace{-15pt}
\end{table}

\subsection{Discussion of Results}
\begin{figure*}[tb]
    \centering
    \includegraphics[width=\textwidth]{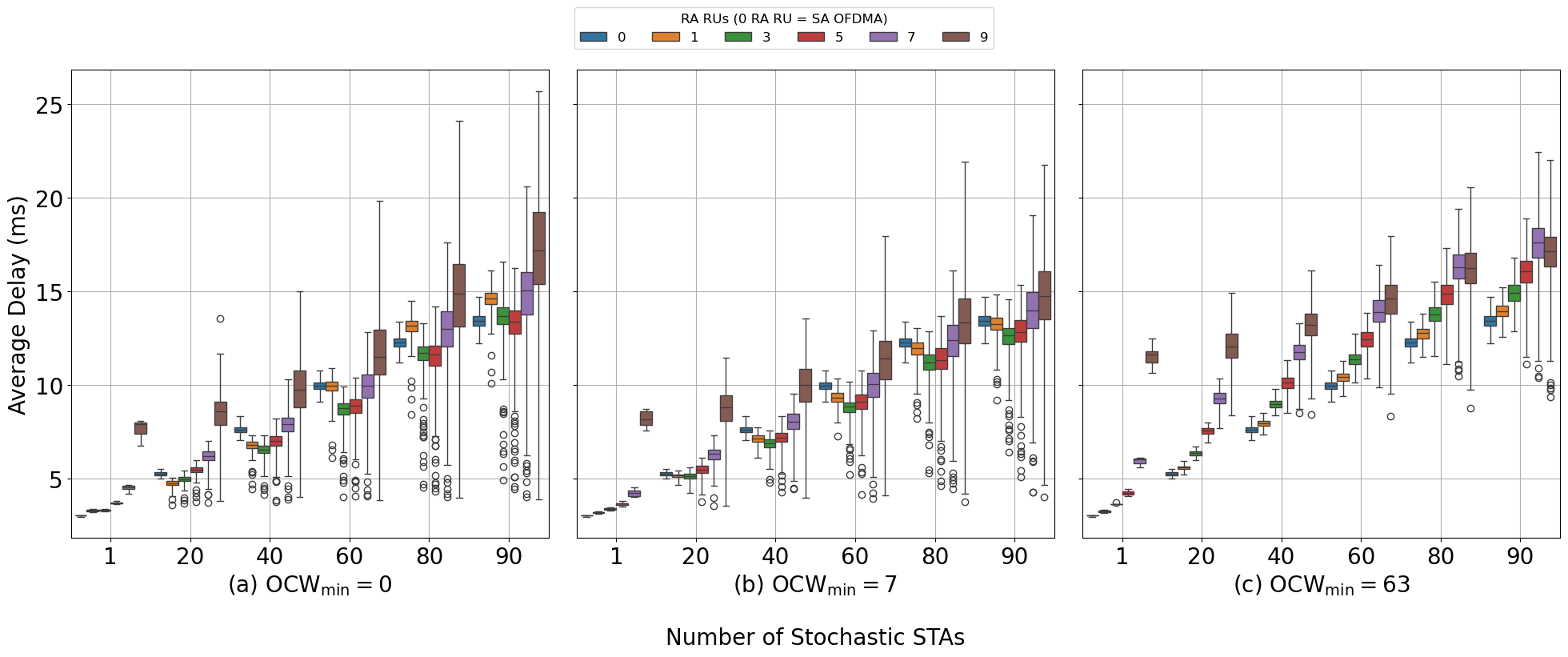}
    \caption{Average delay of stochastic STAs as a function of the number of
    these STAs across different minimum contention window sizes}
    \label{fig:DelayOcwRaRu}
\end{figure*}
\begin{figure}[tb]
    \centering
    \includegraphics[scale=.46]{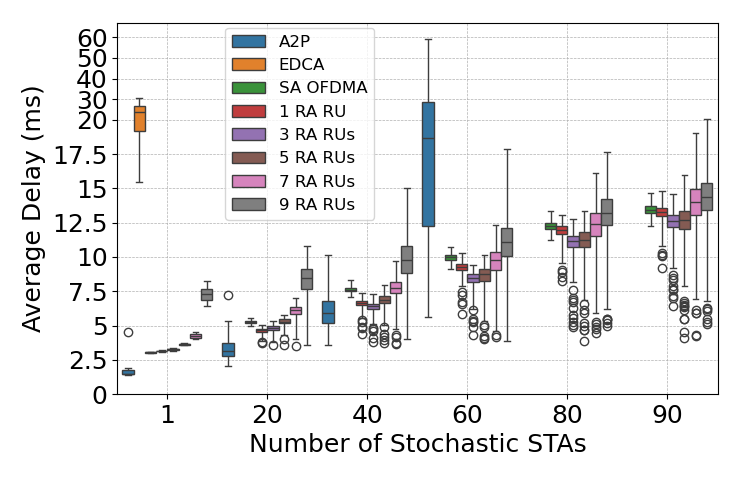}
    \vspace{-.4cm}
    \caption{Average delay of stochastic STAs as a function of the number of
    these STAs}
    \label{fig:all_algo_delay}
    \vspace{-.3cm}
\end{figure}
\begin{figure}[tb]
    \centering
    \includegraphics[scale=.46]{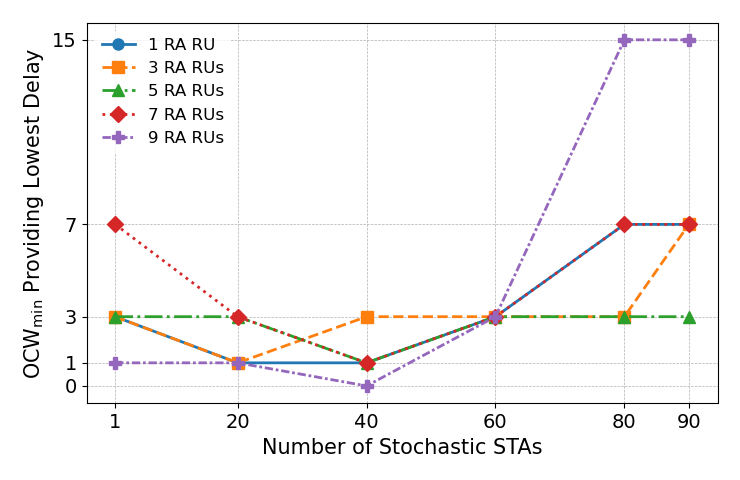}
    \vspace{-.3cm}
    \caption{$\text{OCW}{\min}$ providing the lowest delays as a function of
    the number of stochastic STAs}
    \label{fig:OCWmin_for_MinDelay}
\end{figure}
\begin{figure}[tb]
    \centering
    \includegraphics[scale=.46]{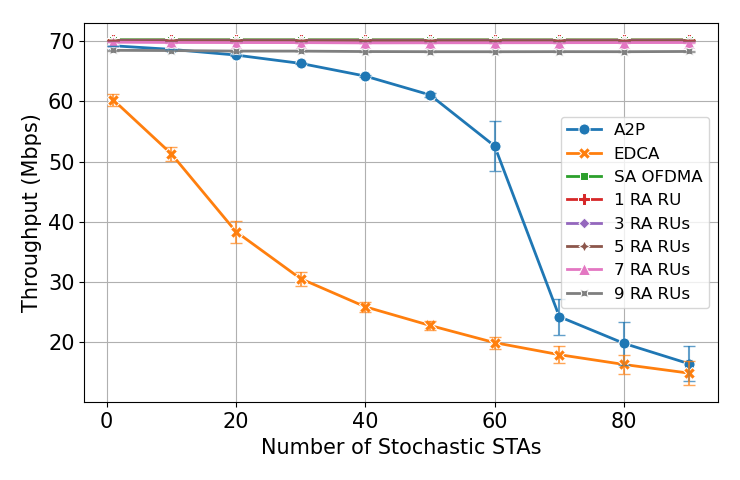}
    \vspace{-.3cm}
    \caption{Total throughput as a function of the number of stochastic STAs}
    \label{fig:all_algo_throughput}
\end{figure}
\begin{figure}[tb]
    \centering
    \includegraphics[scale=.46]{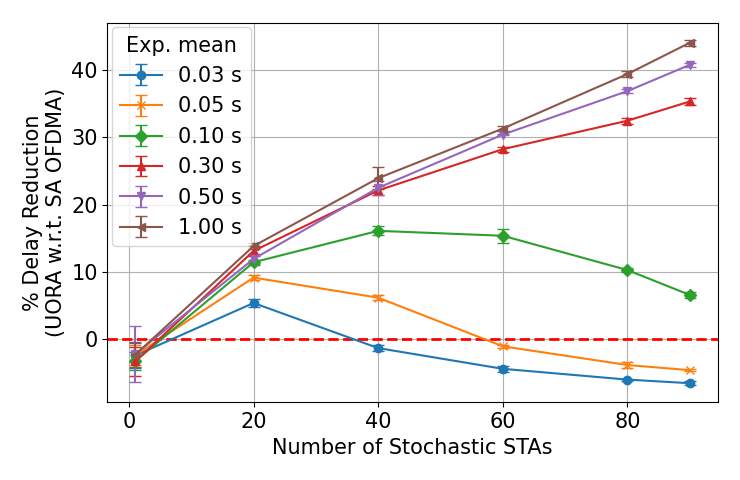}
    \vspace{-.3cm}
    \caption{Delay reduction achieved by UORA with respect to UL OFDMA as a
    function of the number of stochastic STAs for varying intensity of traffic
generated on these STAs. Data points above the dashed red line indicate UORA
outperforming SA OFDMA}
    \label{fig:delayGain}
    \vspace{-.3cm}
\end{figure}

In this section, we compare the performance of UORA with the other schemes in
terms of delay and throughput. We also investigate how the intensity of the
sporadic traffic generated by the stochastic STAs influence the delay reduction
associated with the usage of UORA.

Simulations corresponding to the results presented in
Figures~\ref{fig:DelayOcwRaRu},~\ref{fig:all_algo_delay},
~\ref{fig:OCWmin_for_MinDelay} and~\ref{fig:all_algo_throughput} are conducted
using a packet generation model with an exponential inter-arrival time having a
mean of \qty{100}{\milli\second}. Figures~\ref{fig:DelayOcwRaRu}
and~\ref{fig:all_algo_delay} show the uplink delay for a varying number of
stochastic STAs generating latency-sensitive traffic. The results are
represented in a box plot, where the median, lower, and upper quartiles, and
extreme values of delay are clearly visible---particularly under higher numbers
of stochastic STAs. We begin our analysis by examining the impact of the chosen
$\text{OCW}_\text{min}$ parameter on the performance of UORA. As illustrated in
Figure~\ref{fig:DelayOcwRaRu}, an inappropriate selection of $\text{OCW}{\min}$
can significantly degrade UORA’s performance relative to SA OFDMA (i.e., the
configuration with 0 RA RUs). Conversely, an appropriate choice of
$\text{OCW}_\text{min}$ can yield notable improvements in delay performance.
Specifically, setting $\text{OCW}{\min} = 63$ results in performance inferior to
SA OFDMA, whereas selecting $\text{OCW}{\min} = 7$ leads to reduced average
delay compared to $\text{OCW}{\min} = 0$. Consequently, we select the optimum
$\text{OCW}{\min}$ value to show the best achievable performance of UORA in the
subsequent analysis.

\subsubsection*{\textbf{Average Delay}}
Figure~\ref{fig:all_algo_delay} includes the EDCA result for only
one stochastic STA, as EDCA's performance is already significantly degraded
under this load. For A2P, results are shown for up to 60 STAs, since the delay
increases and throughput decreases substantially beyond this point. Furthermore,
each UORA box plot (i.e., 1-9 RA RUs) represents the distribution of individual
packet delays obtained from the experiment configuration that yields the lowest
average delay across the different $\text{OCW}_\text{min}$ values for a given
number of stochastic STAs $N$ and number of RA RUs $R$. The optimal
$\text{OCW}_\text{min}$ value is determined independently for each $(N, R)$
pair, meaning that the best-performing $\text{OCW}_\text{min}$ setting may vary
across different combinations of $(N, R)$ pairs, as shown in
Figure~\ref{fig:OCWmin_for_MinDelay}. To accomplish this, we conduct $T$
independent experiment runs for every combination of number of stochastic
STAs ($N \in \{1, 10, 20,\dots, 90\}$), number of RA RUs ($R \in \{ 1, 3, 5, 7,
9\}$) and minimum OCW parameter $\text{OCW}_\text{min} \in \{1, 3, 7, 15, 31,
63\}$. For each configuration $(N, R,\text{OCW}_{\min})$, we compute the average
delay across the $T$ runs as:
\[
\bar{D}(N, R, \text{OCW}_{\min}) = \frac{1}{T} \sum_{t=1}^{T} D_t(N, R,
\text{OCW}_{\min}),
\]
where $D_t$ is the average delay observed in the $t^{th}$ run.\\
We then identify, for each $(N, R)$ pair, the $\text{OCW}_\text{min}$ that
results in the lowest average delay:
\[
\text{OCW}^*_\text{min}(N, R) = \arg\min_{\text{OCW}_\text{min}} \bar{D}(N, R,
\text{OCW}_\text{min})
\]
Figure~\ref{fig:OCWmin_for_MinDelay} depicts the results of the optimum
$\text{OCW}_\text{min}$ value for each unique $(N, R)$ pair.
The final box plot for each $(N, R)$ combination in
Figure~\ref{fig:all_algo_delay}, visualizes the distribution of individual
packet delays aggregated from the $T$ runs conducted using that combination's
identified optimal $\text{OCW}^*_{\min}(N, R)$.

For instance, for 40 stochastic STAs and 3 RA RUs, all six $\text{OCW}_{\min}$
settings are evaluated over $T$ independent runs each, and the setting yielding
the lowest average delay is selected. The corresponding box plot then aggregates
the individual packet delays from all $T$ runs conducted using that optimal
$\text{OCW}_\text{min}$ value.

While in EDCA all associated devices (i.e., both deterministic and stochastic
STAs) compete for the channel and are prone to packet losses due to collisions,
the other schemes manage to orchestrate transmissions with little or no
contention. In particular, A2P consistently disables EDCA on deterministic STAs
and, for stochastic STAs, for 8 TUs, thereby reducing contention between STAs
with new packets and the AP. As a result, A2P achieves the lowest delay for up
to 40 stochastic STAs. The sharp increase in delay observed beyond this number of
stochastic STAs is due to heightened contention among devices which prevents
stochastic STAs from getting on the polling list to be subsequently allocated
resources. With the gradual increase of stochastic STAs in the polling list, the
scheduler's ability to efficiently allocate RUs to stochastic STAs, for buffer
status reporting and subsequent data transmission, diminishes. This explains the
gradual increase in delay with a growing number of stochastic STAs observed in
the contention-free SA OFDMA scenario. Moreover, contention in UORA is less
aggressive than in EDCA, as it utilizes multiple RUs. The aggressiveness of UORA
contention depends on the values of $\text{OCW}_\text{min}$ and the number of
RA-RUs, both of which can be dynamically adjusted based on network load. As
visualized in Figure~\ref{fig:OCWmin_for_MinDelay}, different traffic
loads---characterized by the number of stochastic STAs---and varying
allocations of RA RUs require distinct contention window sizes in order to
achieve optimal UORA performance. Both SA OFDMA and UORA emerge as the most
scalable solutions, maintaining delay below \qty{15}{\milli\second} while
supporting up to 90 stochastic STAs. 

The results further show that increasing the number of RUs reserved for random
access reduces delay as the population of latency-sensitive devices grows---up
to the point where RUs are equally divided between scheduled access (SA) and RA
(i.e., 5 RA RUs). Beyond this midpoint of equal RA and SA allocation, the
delay begins to increase again. This behavior can be attributed to the trade-off
between contention-based (UORA) and contention-free (SA OFMDA) access
mechanisms. As more RUs are allocated for RA, stochastic STAs benefit from
increased transmission opportunities by not having to wait to be scheduled,
which helps reduce delay, particularly when the number of such devices is high.
However, once the RA allocation surpasses the point of balance with SA, more
stochastic STAs begin to rely on UORA, which exacerbates contention and
collision rates, ultimately increasing the delay experienced by packets.

\subsubsection*{\textbf{Throughput}}
As discussed in Section~\ref{sec:method}, with SA OFDMA and UORA, resource
allocation for BSRs from STAs is decoupled from RUs assignment for data
transmission. As such, RUs in the BSRP TF that experience collisions during BSR
transmission can be rescheduled in Basic TF for uplink data transmission. In our
simulations, the AP reserves RUs for RA only in the BSRP TF, while all RUs in
the Basic TF are assigned for SA uplink data transmission. Consequently, the
throughput remains fairly constant throughout the experiment, as shown in
Figure~\ref{fig:all_algo_throughput}. The relatively lower throughput observed
when all RUs are reserved for RA is attributed to the contention-based mechanism
used to select an RU for BSR transmission. This mechanism can lead to resource
underutilization, as the AP may fail to receive BSRs from multiple STAs,
including those with deterministic traffic patterns.

\subsubsection*{\textbf{Delay Reduction}}
We further investigate the variations in delay between SA OFDMA transmissions
(i.e., RA RU = 0) and UORA-based (i.e., RA RUs $> 0$) transmissions under
varying traffic intensities generated by critical STAs. The objective is to
study how the delay reduction that can be obtained by using UORA varies with the
intensity of latency-sensitive traffic required for UORA to provide measurable
performance benefits. Figure~\ref{fig:delayGain} depicts the percentage of delay
reduction achieved with UORA relative to the baseline case of SA OFDMA across
different values of the exponential mean used in the traffic generation model.
Values above \num{0} indicate that UORA outperforms SA OFDMA, whereas values
below \num{0} indicate the opposite. We compute the delay reduction as
follows:

\begingroup
\footnotesize
\[
    D_{gain} = \frac{D_{base} - D_{min}}{D_{base}} \times 100,
    \label{eq:gain}
\]
\endgroup
where $ D_{min}$ denotes the minimum delay observed across the different values
of $\text{OCW}_\text{min}$ and number of RA RUs for each configuration of
exponential mean and number of critical STAs when using RA RUs. The delay when
no RA RUs are used is denoted by $D_{base}$. Additionally, the error bars
represent the standard deviation across the $T$ experimental runs.

The results show that reserving RUs for RA becomes less advantageous when a
large number of critical STAs are actively and frequently generating data. In
such scenarios, the contention-based nature of UORA leads to increased
collisions, thereby diminishing its performance benefits. Conversely, UORA is
more effective under sparse traffic conditions. For example, under relatively
higher traffic loads—characterized by exponential means of \qty{0.30}{\second}
and \qty{0.50}{\second}—the performance gains of UORA begin to decline beyond 40
and 60 critical STAs, respectively. Furthermore, the advantages of using UORA
tend to plateau when the traffic becomes sufficiently sparse, as observed with
an exponential mean of \qty{0.30}{\second} and higher.

\section{Conclusion \label{conclusion}}
In this work, we evaluate the efficacy of UORA as a scalable alternative to
buffer status polling mechanisms such as A2P and SA UL OFDMA. EDCA serves as a
baseline scheme. We demonstrate that UORA effectively mitigates the limitations
of both the contention-based channel access (EDCA) and centralized polling
(SA UL OFDMA), while also offering advantages over hybrid approaches like A2P,
particularly in dense environments with sparsely sporadic uplink traffic. The
results show that UORA  achieves lower delay and higher throughput compared to
the alternative approaches in scenarios with a large number of stochastic STAs
generating latency-sensitive traffic, underscoring its potential to enhance
UL performance in Wi-Fi networks.

In our future work, we plan to investigate ways for the optimal selection of
UORA parameters, e.g., using mathematical optimization.

\section*{Acknowledgements}
The research was supported by the FWO WaveVR project (G034322N).
\printbibliography[title={References}]
\end{document}